# Compaction and condensation of DNA mediated by the C-terminal domain of Hfq

Antoine Malabirade [1,†], Kai Jiang [2,†], Krzysztof Kubiak [1,3], Alvaro Diaz-Mendoza [4], Fan Liu [2], Jeroen A. van Kan [2], Jean-François Berret [4], Véronique Arluison [1,4,*], and Johan R.C. van der Maarel [2,*]

[1]Laboratoire Léon Brillouin, CEA, CNRS, Université Paris Saclay, 91191 Gif-sur-Yvette, France; [2]Department of Physics, National University of Singapore, 2 Science Drive 3, Singapore 117542; [3]Department of Molecular Biology, University of Gdansk, Wita Stwosza 59, 80-308 Gdansk, Poland; [4]Université Paris Diderot, Sorbonne Paris Cité, 75013 Paris, France.

## ABSTRACT

Hfq is a bacterial protein that is involved in several aspects of nucleic acids metabolism. It has been described as one of the nucleoid associated proteins shaping the bacterial chromosome, although it is better known to influence translation and turnover of cellular RNAs. Here, we explore the role of *Escherichia coli* Hfq's C-terminal domain in the compaction of double stranded DNA. Various experimental methodologies, including fluorescence microscopy imaging of single DNA molecules confined inside nanofluidic channels, atomic force microscopy, isothermal titration microcalorimetry, and electrophoretic mobility assays have been used to follow the assembly of the C-terminal and N-terminal regions of Hfq on DNA. Results highlight the role of Hfq's C-terminal arms in DNA binding, change in mechanical properties of the double helix and compaction of DNA into a condensed form. The propensity for bridging and compaction of DNA by the C-terminal domain might be related to aggregation of bound protein and may have implications for protein binding related gene regulation.

## INTRODUCTION

Hfq is a phylogenetically conserved protein found at high concentration in approximately half of the sequenced bacterial genomes (1). Consistent with this abundance, the protein coordinates multiple roles in bacterial metabolism. For instance, the first description of Hfq is related to its role in bacteriophage Qβ RNA replication [Hfq stands for host factor for phage Qβ] (2). The pleiotropic functions of Hfq were highlighted when the *hfq* gene was disrupted in *Escherichia coli*, resulting in various phenotypes (3). Most of these *hfq*-null phenotypes are due to the role of Hfq in RNA regulation. Indeed, even if its requirement seems to be facultative in some bacteria, it is usually needed to mediate post-transcriptional stress-response with small noncoding RNA (sRNA). This regulatory mechanism is based on the hybridisation of sRNA to its target mRNA. Accordingly, the majority of known sRNA interacts nearby the ribosome binding site to prevent initiation of translation and thus favours mRNA decay (4, 5, 6). Few studies have shed light on the importance of Hfq in DNA metabolism. It has been suggested that Hfq affects plasmid replication and DNA transposition (7, 8). Hfq might also be involved in transcription control, such as regulation of transcription initiation and/or elongation (9, 10). Hfq might also regulate transcription by protein interaction, mainly with RNA polymerase and Rho transcription termination factor (11, 12).

Hfq has been described as one of the nucleoid associated proteins (NAPs) shaping the bacterial chromosome (13). The amount of Hfq in the bacterial cell is comparable to those of the most abundant NAPs, that is Fis (factor for inversion simulation) and HU (heat-unstable protein) (14). Its quantity in the nucleoid might however be less than the ones for HU and Fis, as only 10-20% has been estimated to be chromosome bound (15, 16). The cellular concentration of Hfq increases when reaching the stationary phase, but the nucleoid-bound fraction remains approximately constant (7, 16, 17, 18). Hfq has a significant affinity for DNA with an equilibrium dissociation constant $K_d$ in the range from nano- to micromolar (the values of $K_d$ pertaining to binding to RNA are three orders of magnitude smaller) (10, 19). Hfq not only interacts with DNA but also with other NAPs such as H-NS (20, 21). Hfq belongs to a family of NAPs that form bridges between sections of the DNA molecule, thereby organising large parts of the chromosome into domains (10, 22). It has also been suggested that Hfq constrains negative supercoiling *in vivo* (3). However, the mechanism how Hfq affects DNA topology is unclear.

Structurally, the 102 amino acids Hfq from *Escherichia coli* forms an Sm-fold in its N-terminal region (NTR, 65 amino acids) (23). The NTR consists of a strongly bent, five stranded antiparallel $\beta$-sheet capped by an $\alpha$-helix. The $\beta$-sheets from six monomers assemble into a toroidal structure with two faces, that is a proximal (where the $\alpha$-helix is located) and a positively charged, distal face (23). The distal face, the edge, and the C-terminal region (CTR, 35 amino acids) are involved in DNA binding. This information is based on experiments with hfq mutants with deleted CTR or a variety of point mutations such as K31A, Y25A (distal face) or R16A (edge) (24). While the CTR is non-essential for RNA binding, the distal face and edge are involved in both DNA and RNA fixation (25). *In vitro*, the protein shows a preference for A-tracts ($K_d$ = 200 nM) (10). A-tracts are usually associated with an intrinsic curvature of the DNA molecule (13). *In vivo*, Hfq has a preferred (A/T)T(A/G)TGCCG binding motif with a curved topology (24). Besides these preferred binding sequences, Hfq also interacts with DNA in a sequence-nonspecific manner (13, 26). Although the CTR region is usually considered to be intrinsically disordered, it was recently reported that it contains an amyloid sequence that allows the protein to self-assemble (4, 27).

In previous work, we have focussed on DNA-compacting properties of wild-type Hfq (28). Various experimental technologies, including fluorescence microscopy imaging of single DNA molecules confined inside nanofluidic channels, atomic force microscopy and small angle neutron scattering were used to follow the assembly of Hfq on DNA. It was shown that Hfq forms a nucleoprotein complex, changes the mechanical properties of the double helix and compacts DNA into a condensed form. A compaction mechanism based on protein-mediated bridging of DNA segments was proposed. The propensity for bridging was hypothesised to be related to the multi-arm functionality of the Hfq hexamer, resulting from binding of the C-terminal regions to the duplex. These previous experiments were done with full length Hfq. Here, we report similar molecular imaging experiments, electrophoretic mobility assays (EMSA) of DNA in the presence of a 38 amino acids peptide representing the C-terminal region (Hfq–CTR) and a truncated version of the protein composed of the 72 first amino acids, representing the N-terminal region (Hfq–NTR). Note that Hfq–NTR forms the hexameric torus, whereas Hfq–CTR represents the protruding C-terminal arms of wild type Hfq. The thermodynamics of Hfq–CTR's association with DNA is investigated using isothermal titration calorimetry (ITC). Our new results enable to firmly establish the role of the C-terminal domain in Hfq-mediated bridging and compaction of double stranded DNA.

## MATERIALS AND METHODS

### Sample preparation

His-tagged *Escherichia coli* Hfq–NTR (residues 1-72) was purified from over expressing BL21-(DE3)$\Delta$*hfq*/pLATE11-*hfqntr* cells. The pLATE11-*hfqntr* expression vector was constructed by the ligation independent cloning method (Thermo Fisher). The oligonucleotides used for cloning were AGAAGG-AGATATAACTATGGCTAAGGGGCAATCTTTACAA-GATC (for) and GGAGATGGGAAGTCATTAACTGT-GATGAGAAACCGGGC (rev). Insertion of the His-tag was performed by site-directed mutagenesis (Q5, New England Biolabs). As the His-tag potentially modifies the properties of Hfq, a Tobacco Etch Virus (TEV) protease site (ENLYFQG) was inserted between the N-terminal His-tag and Hfq sequence. The oligonucleotides used for this double insertion were GAGAACCTGTACTTCCAGGGCGCTAAGGGGCAA-TCTTTAC (for) and ACCATGATGATGATGATGATG-CATAGTTATATCTCCTTCTGG (rev). Cells from post-induction cultures were resuspended in 20 mM Tris–HCl *p*H 7.5, 0.5 M NaCl, 10% (v/v) glycerol and a protease inhibitor (Sigma) at 277 K. The suspension was sonicated and the lysed cells were cleared by centrifugation at 15,000 g for 30 min. DNase I (40 g/L) and RNase A (30 g/L) were added to the cleared lysate at 303 K. The solution was then applied to a $Ni^{2+}$–NTA column (GE Healthcare). The resin was washed with 20 mM Tris–HCl, *p*H 7.8, 0.3 M NaCl, 20 mM imidazole and the protein was eluted with a gradient of imidazole (20–500 mM). TEV digestion was carried out according to the manufacturer's instructions (Thermo Fisher). After digestion, Hfq–NTR was purified with a HiTrap Q HP 1 mL column (GE Healthcare). Before injection on the anion exchange column, Hfq–NTR was diluted five times in equilibration buffer (50 mM $NaH_2PO_4$, *p*H 7) to reduce salt concentration. The resin was equilibrated with the same buffer and the protein was eluted with a gradient of NaCl (0-1 M). Digested His-tag, non-digested Hfq–NTR and contaminants were removed with the anion exchange column (27). Hfq–CTR (residues 64-102) was obtained from Genosphere Biotech (France). Full length Hfq has been prepared and purified as described before (28).

T4 GT7 DNA (T4–DNA, 165.65 kbp) was obtained from Nippon Gene, Tokyo and used without further purification. No DNA fragments of ones to tens of kbps were observed by pulsed gel electrophoresis. Samples were prepared by dialysing solutions of DNA against 10 mM Tris-HCl (T) buffer and T–buffer with 30 mM potassium glutamate (KGlu). Solutions of Hfq–CTR and Hfq–NTR in the same buffer were also prepared. The Tris-HCl concentration is 10 mM Tris adjusted with HCl to *p*H 7.5 (that is 8.1 mM TrisCl and 1.9 mM Tris). The ionic strength of the buffer was calculated with the Davies equation for estimating the activity coefficients of the ions and a dissociation constant *p*K = 8.08 for Tris. Solutions of Hfq–CTR or Hfq–NTR and DNA were mixed and incubated for 12 h at 277 K. YOYO-1 fluorescence staining dye was purchased from Invitrogen, Carlsbad, CA. DNA was stained with YOYO-1 with an incubation time of 12 h and an intercalation ratio of 100 bps per dye. No anti-photo bleaching agent was used.

### Electrophoretic mobility shift assay (EMSA)

The binding of Hfq, Hfq–CTR, and Hfq–NTR to DNA was investigated through the measurement of the equilibrium dissociation constant ($K_d$) with a gel shift assay. A 1,000 bp DNA fragment was obtained from Thermo Scientific. The fragments (fragment concentration of 15 nM) were incubated with an excess of Hfq, Hfq–CTR, and Hfq–NTR, respectively, at room temperature for 20 min in T–buffer with 50 mM NaCl, $p$H 7.5. Band shifts were resolved on non-denaturing gradient 4-12% polyacrylamide gel. The native gel was run for 2 h at room temperature with 40 mM Tris-Acetate, 1 mM EDTA, $p$H 8.0 (TAE) buffer, stained with GelRed (Biotium), and imaged with a G:BOX system (Syngene, Cambridge, UK). Quantification of bands was achieved by IMAGEJ software (http://rsb.info.nih.gov/ij/). Depending on the protein, either a cooperative (sigmoidal) or non-cooperative (hyperbola) binding model was applied. Affinities have been estimated through the measurement of the equilibrium dissociation constant $K_d$ in a non-cooperative model (expressed in mol/L). In case of the cooperative model, the apparent dissociation constant $K_d^{ap}$ is reported together with the Hill coefficient $n_h$ as an indication of the extent of the cooperativity.

### Chip fabrication

The channel systems were fabricated by replication in PDMS of patterned master stamps (29, 30). The nanochannel part of the stamps was made in HSQ resist (Dow Corning, Midland, MI) using a lithography process with proton beam writing (31). An array of nanochannels is connected to two loading reservoirs through a superposing set of microchannels made in SU-8 resin with UV lithography. The heights and widths of the ridges in the master stamps were measured with atomic force microscopy (Dimension 3000, Veeco, Woodbury, NY) and scanning electron microscopy, respectively. Two stamps were made featuring nanochannels of length 60 $\mu$m and rectangular cross-sections of $150 \times 250$ and $200 \times 300$ nm$^2$, respectively. The stamp was coated with a 5 nm thick teflon layer for perfect release of the replicated PDMS chips (32). The stamps were replicated in PDMS followed by curing with a curing agent (Sylgard, Dow Corning) at 338 K for 24 h. The replicas were sealed with a glass coverslip, after both substrates were plasma oxidised (Harrick, Ossining, NY).

### Nanofluidics

The pre-incubated and stained DNA molecules were loaded into one of the two reservoirs of biochip. The DNA molecules were subsequently driven into the nanochannels by electrophoresis. For this purpose, two platinum electrodes were immersed in the reservoirs and connected to a power supply with a voltage in the range 0.1–10 V (Keithley, Cleveland, OH). Once the DNA molecules were localised inside the nanochannels, the electric field was switched off and the molecules were allowed to relax to their equilibrium state for at least 60 s. The stained DNA molecules were visualised with a Nikon Eclipse Ti inverted fluorescence microscope equipped with a 200 W metal halide lamp, a filter set, and a 100× oil immersion objective. A UV light shutter controlled the exposure time. Images were collected with an electron multiplying charge coupled device (EMCCD) camera (iXon X3, Andor Technology, Belfast, UK). One minute video clips were collected with a frame rate of 5 frames per second. The video clips were analysed with MATLAB (Natick, MA). For each frame (time $t$), the radius of gyration tensor of the imaged molecule was constructed according to $\mathbf{S}(t) = \sum_{m,n} (\mathbf{r}_{mn} - \mathbf{r}_{cm})^2 I_{mn}/I_0$, with $I_{mn}$ the fluorescence intensity of pixel $[m,n]$ at position $\mathbf{r}_{mn}$ in the $xy$–plane and $I_0$ is the total intensity of the frame. For intensity threshold, we have used two times the signal to background noise ratio. The centre of mass $\mathbf{r}_{cm}$ of the molecule was calculated according to $\mathbf{r}_{cm} = \sum_{m,n} \mathbf{r}_{mn} I_{mn}/I_0$. We determined the principal eigenvalues and vectors of $\mathbf{S}$ using a singular value decomposition and with the assumption of cylindrical symmetry. We subsequently derived the extension of the molecule along the direction of the channel according to $L_{\parallel}^2 = 12 < S_{\parallel} >$, where $S_{\parallel}$ represents the largest eigenvalue and the brackets denote an average over all frames.

### Bulk phase imaging

A droplet of solution was deposited on a microscope slide and sealed with a coverslip separated by a 0.12 mm spacer. The YOYO-1 stained T4–DNA molecules were imaged with the above mentioned microscope. One minute video clips were collected with a frame rate of 5 frames per second. The video clips were analysed as described above, but with the largest eigenvalue of the radius of gyration tensor corresponding to the length of the long axis of the molecule.

### Atomic force microscopy

DNA fragments were purchased from Thermo Scientific (1,000 and 10,000 bp, Waltham, MA). All imaging experiments were done at room temperature in air with a Veeco Dimension 3000 atomic force microscope (Woodbury, NY). Images were acquired in the tapping mode with silicon (Si) cantilevers (spring constant of 20–100 N/m) and operated below their resonance frequency (typically 230–410 kHz). The images were flattened, and the contrast and brightness were adjusted for optimal viewing conditions. A 5 $\mu$L droplet was spotted onto a silica surface. After 10 min to allow for DNA adsorption onto the surface, the specimens were developed by flushing them with ultra pure water followed by drying in a stream of N$_2$ gas.

### Isothermal Titration Calorimetry

Isothermal titration calorimetry (ITC) was performed using a Microcal VP-ITC calorimeter (Northampton, MA) with a cell volume of 1.464 mL, working at 298 K and an agitation speed of 307 rpm. The syringe and the measuring cell were filled with degassed solutions of DNA (1,000 bp) and Hfq–CTR in T–buffer ($p$H 7.5), respectively. Typical concentrations in the syringe were between 1 and 4 mmol of DNA bps/L, whereas the measuring cell contains the Hfq–CTR peptide dispersion at molar concentrations of 0.05 to 0.5 mM. All experiments were done by adding DNA to the peptide. The titration experiment consisted of a preliminary injection of 2 $\mu$L, followed by 28 injections of 10 $\mu$L at 10 min intervals. A typical ITC experiment consists in measuring the differential power provided by the calorimeter to keep the temperature of the cell constant after each injection. In a second step, the integration over time of the ITC signal was performed and translated in terms of binding enthalpy as a function of the molar ratio $X = [\text{DNA}]/[\text{Hfq–CTR}]$. Control experiments corresponding to dilution of DNA with T–buffer were also done and the corresponding integrated signal was subtracted from the raw binding enthalpy.

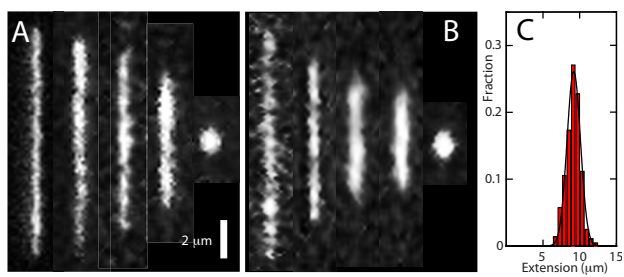

**Figure 1.** **(A)** Montage of fluorescence images of T4–DNA molecules inside $200 \times 300$ nm$^2$ channels and in T–buffer (pH 7.5). From left to right $1 \times 10^{-4}$, $1 \times 10^{-3}$, $1 \times 10^{-2}$, $5 \times 10^{-2}$, and $1 \times 10^{-1}$ (condensed) $\mu$M Hfq–CTR. **(B)** As in panel A, but for T4–DNA inside $150 \times 250$ nm$^2$ channels and in T–buffer with 30 mM KGlu. **(C)** Distribution in extension of a population of 30 molecules in T–buffer with $1 \times 10^{-3}$ $\mu$M Hfq–CTR. A Gaussian fit gives a mean extension $R_\parallel = 9 \pm 1$ $\mu$m.

## RESULTS

### Nanofluidics

Prior to the nanofluidic experiments, T4–DNA molecules with a concentration of 3 mg of DNA/L were incubated with the relevant buffer for at least 24 h. The buffers are 10 mM Tris/HCl (T–buffer) and T–buffer with 30 mM KGlu (pH 7.5). The DNA molecules were subsequently driven into the channels of the nanofluidic device by electrophoresis. We have used two different devices with channel cross-sections of $200 \times 300$ and $150 \times 250$ nm$^2$, respectively. After switching off the electric field, the stretch of the molecules along the longitudinal direction of the channels equilibrates within 60 s. Montages of fluorescence microscopy images showing equilibrated, single T4–DNA molecules in the presence of various concentrations of the C-terminal domain (Hfq–CTR) pertaining to the two channel systems are shown in panels A and B of Figure 1, respectively. With increasing concentration of Hfq–CTR, the stretch decreases and, eventually, the DNA molecule compacts into a condensed form. Condensed DNA is clearly discernible as a bright fluorescence spot. In the case of the N-terminal torus (Hfq–NTR), the stretch shows a moderate decrease with increasing concentration of the protein, but no condensation is observed (montages for Hfq–NTR are not shown).

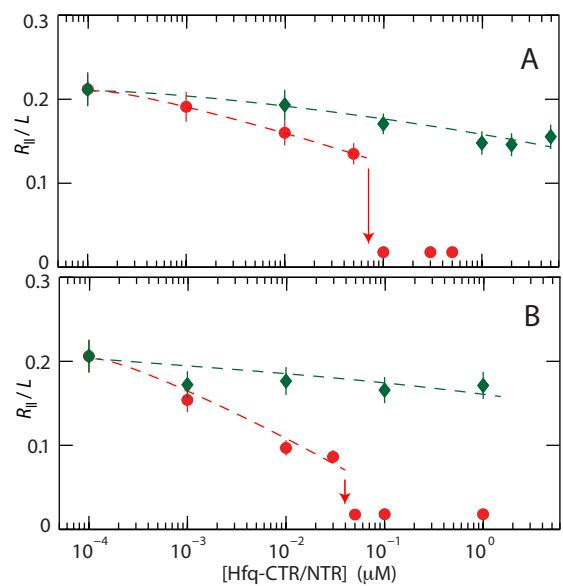

**Figure 2.** **(A)** Relative extension $R_\parallel/L$ of T4–DNA inside $200 \times 300$ nm$^2$ channels and in T–buffer versus the concentration of Hfq–CTR (red, ○) and Hfq–NTR (green, ◊). **(B)** As in panel A, but for T4–DNA inside $150 \times 250$ nm$^2$ channels and in T–buffer with 30 mM KGlu. The dashed curves are drawn as an aid to the eye and the arrows denote the condensation thresholds.

For each channel system and buffer composition, we measured the stretch of about 30 DNA molecules in a fresh PDMS replica. The observed distribution in length is close to Gaussian (an example is shown in panel C of Figure 1). Fragmented DNAs with an extension less than the mean value minus two times the standard deviation were discarded. The mean relative extension $R_\parallel/L$, that is the mean stretch divided by the contour length of the T4–DNA molecule (57 $\mu$m), is set out in Figure 2 versus the concentration of the relevant protein. Note that the protein concentrations refer to the C– and N–terminal domains of Hfq in monomeric form. With increasing concentration of Hfq–CTR, the relative extension decreases. For an over-threshold concentration of Hfq–CTR, the DNA molecules compact into a condensed form. A small decrease in stretch and no condensation are observed by increasing concentration of Hfq–NTR. Qualitatively, the same behaviour is observed for the two channel systems. The increase in stretch caused by stronger confinement in a channel of smaller cross-sectional diameter offsets the decrease in stretch caused by the increase in concentration of salt.

For over-threshold concentrations of Hfq–CTR, the DNA molecules compact into a condensed form. Notice that the condensation is facilitated by nano-confinement. In the feeding microchannels and reservoirs of the chip, the DNA molecules are in the non-condensed, coiled form. The critical concentration of Hfq–CTR for condensation inside $150 \times 250$ nm$^2$ channels and in T–buffer with 30 mM KGlu is $0.07 \pm 0.02$ $\mu$M. For the wider $200 \times 300$ nm$^2$ channel system and lower ionic strength (T–buffer), this critical concentration decreases somewhat to 0.04±0.01 $\mu$M. With a bp concentration of 4.6 $\mu$M, the DNA molecule hence condenses with an Hfq–CTR to bp ratio on the order of 1:100. In the case of the N-terminal torus Hfq–NTR, no condensation is observed.

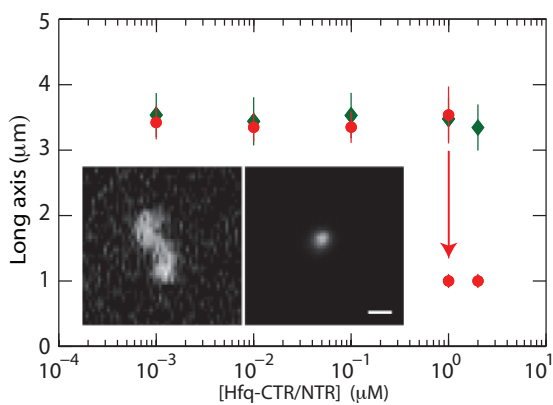

**Figure 3.** Long axis of unconstrained T4–DNA in T–buffer versus the concentration of Hfq–CTR (red, ○) and Hfq–NTR (green, ◇). The dashed curves are drawn as an aid to the eye and the arrow denotes the condensation threshold (for Hfq–CTR only). The inset shows fluorescence images of a non-condensed (left) and condensed (right) T4–DNA molecule.

### Condensation in the bulk phase

The size of unconstrained DNA coils in the presence of various concentrations of Hfq–NTR or Hfq–CTR was investigated with fluorescence microscopy. The experiments were done with 0.03 mg of T4–DNA/L in T–buffer (pH 7.5). Notice that the concentration of DNA is two orders of magnitude lower than the one pertaining to the loading buffer in the nanofluidics experiments. Furthermore, the DNA molecules were stained with a minimal level of staining of 100 bps per YOYO-1 molecule. Such a low staining level ensures no appreciable effect on the contour length, net charge and bending rigidity of the DNA molecules (29, 33). Unconstrained DNA coils are typically slightly anisotropic and their physical extent is best represented by the length of the long axis of the radius of gyration tensor. The measured lengths of the long axis, as well as some characteristic fluorescence images, are shown in Figure 3.

The length of the long axis pertaining to non-condensed molecules is about 3.5 $\mu$m and almost constant with increasing concentration of the peptides. For over-threshold concentrations of Hfq–CTR, the DNA molecules compact into a condensed form. A difference is the coexistence of condensed and non-condensed molecules at the critical concentration of Hfq–CTR in the bulk phase, indicating a discontinuous, first order transition. In the channel systems, the transition appears to be continuous (34). The critical concentration of Hfq–CTR for condensation is 1.0 $\mu$M. Condensation in the bulk phase hence requires more than an order of magnitude higher concentration of Hfq–CTR. In the case of Hfq–NTR, no condensation in the bulk phase is observed.

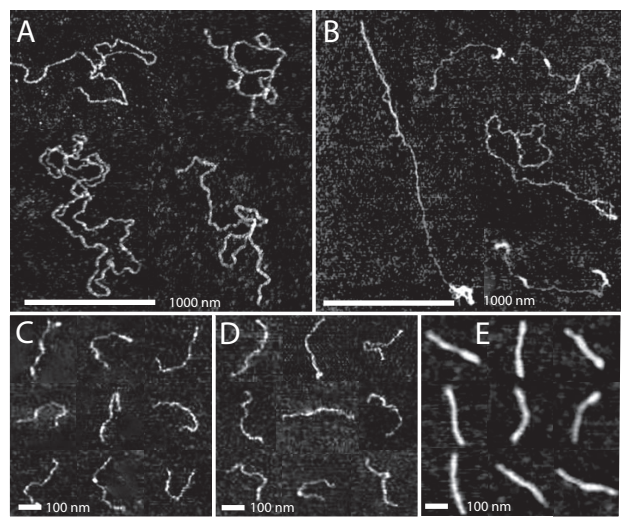

**Figure 4.** (**A**) Tapping mode atomic force microscopy images of DNA (10,000 bps) with Hfq–CTR to bp ratio of 1:50. (**B**) As in panel A, but with Hfq–CTR to bp ratio of 1:1. (**C**) DNA (1,000 bp) with Hfq–CTR to bp ratio of 1:50. (**D**) As in panel C, but with Hfq–CTR to bp ratio of 1:1. (**E**) As in panel C, but with Hfq–CTR to bp ratio of 10:1. Panels A-E are a montage.

### Contour and persistence length

The nanofluidic and bulk phase results indicate interaction of Hfq–CTR with DNA, which eventually results in compaction of the DNA molecule into a condensed form. We have investigated the length of the DNA molecule along its contour as well as its persistence length in the presence of Hfq–CTR with atomic force microscopy. Linear DNA molecules of 10,000 and 1,000 bps were used at a concentration of 3 mg/L. The molecules were incubated with 0.1, 5, and 50 (1,000 bp DNA only) $\mu$M Hfq–CTR for 12 h, which correspond with an Hfq–CTR to bp ratio of about 1:50, 1:1, and 10:1 respectively. In the presence of Hfq–CTR, no additions to the buffer are necessary to promote adhesion of the DNA molecules to silica. The weakly adsorbed molecules equilibrate on the surface in a 2D conformation. Montages of images for 10,000 bp DNA with increasing Hfq–CTR to bp ratio are shown in panels A and B of Figure 4. The results for 1,000 bp DNA are shown in panels C through E. There is no adhesion of DNA to the silica surface without peptide or protein and/or in the presence of Hfq–NTR.

For 10,000 bp DNA with an Hfq–CTR to bp ratio of 1:50 (Figure 4A) the molecules are spread on the surface without obvious side-by-side aggregation. However, for a larger Hfq–CTR to bp ratio of 1:1, intramolecular back folding and looping is observed (Figure 4B). The 1,000 bps molecules are visible as semi-flexible rods and do not exhibit aggregation nor looping. In the presence of Hfq–CTR there is no obvious change in the averaged heights of the molecules (0.9±0.1 nm). The measured values of the height are however indicative, because the complexes are dried and spread on the silica surface. To obtain the contour length, the centreline of the molecules was traced. The distributions in length are displayed in panels A-C of Figure 5 and the averaged values and standard deviations are collected in Table 1. For an Hfq–CTR to bp ratio of 1:50 and 1:1 the average value of the contour length of agrees with the contour length of 1,000 bps DNA in the B-form (340 nm). For a relatively large Hfq–CTR to bp ratio of 10:1, the contour length takes a value of 285±20 nm. The latter value is in good agreement with the contour length pertaining to the A-form (290 nm).

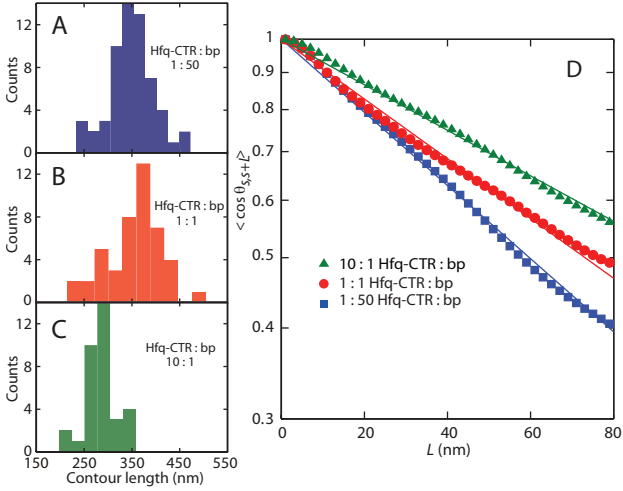

**Figure 5.** (**A–C**) Distribution in contour length of DNA (1,000 bp) molecules with Hfq–CTR to bp ratio of 1:50 (blue), 1:1 (red) and 10:1 (green). (**D**) Orientation correlation function of the tangent vectors at a pair of points separated by distance $L$ along the contour. The closed symbols are the experimental data obtained by averaging 34–49 molecules. The solid lines are exponential fits.

**Table 1.** Contour length $L$ and persistence length $P$ of 1,000 bp DNA for various Hfq–CTR to bp ratios as determined by AFM.

| Hfq–CTR : bp | $L$ (nm)   | $P$ (nm)       |
|--------------|------------|----------------|
| 1:50         | 340 ± 20   | 43.1 ± 0.3     |
| 1:1          | 350 ± 20   | 52.7 ± 0.4     |
| 10:1         | 285 ± 20   | 69.2 ± 0.6     |

The centrelines were also used to obtain the tangent vector correlation function $\langle \cos\theta_{s,s+L} \rangle$, where $\theta$ is the angle between tangent vectors at points $s$ and $s+L$, by averaging $s$ along the contour (35). For DNA molecules equilibrated in 2D conformation, the correlation functions takes the form $\langle \cos\theta_{s,s+L} \rangle = \exp[-L/(2P)]$. The inverse of the exponential decay constant gives hence the persistence length $P$. Experimental and fitted tangent correlation functions are shown in panel D of Figure 5. The fitted values of $P$ are also collected in Table 1. With increasing Hfq–CTR to bp ratio from 1:50 to 10:1, a moderate increase in the values of $P$ is observed. AFM of naked DNA adsorbed on mica mediated by $Mg^{2+}$ has previously given $P = 56 \pm 4$ nm (36). Accordingly, DNA's flexibility is moderately affected by the interaction with Hfq–CTR.

## DNA binding properties of Hfq protein and domains

The interactions of Hfq, Hfq–CTR and Hfq–NTR with DNA have been explored with EMSA. The gels and band shifts are shown in Figure 6. The concentrations are expressed in moles of monomeric peptide/L. For full length Hfq, a two-step band shift is observed. At lower concentrations of Hfq, the band shifts midway the well and the band pertaining to protein-free DNA. This shift corresponds with a binding mode of higher affinity (apparent dissociation constant $K_d^{ap} = 1.5 \pm 0.3$ μM). When the concentration is increased beyond 1 μM, the intermediate band disappears and is shifted to a migration distance close to the loading wells. This second band shift agrees with a cooperative binding mode of lower affinity (apparent $K_d^{ap} = 6.5 \pm 1.0$ μM) and a Hill coefficient $n_h$ of 3.9±0.3. In the case of Hfq–NTR, a single diffuse band shift is observed but no significant amount of material shifted close to the wells. Several gels were analysed and either no or small cooperativity was observed (Hill coefficient $n_h = 1$–2, apparent $K_d^{ap}$ around 20 μM). For Hfq–CTR, a single band shifted to a migration distance close to the wells is observed for relatively high concentrations of the peptide. The latter band shift agrees with a binding mode of relatively low affinity ($K_d = 240 \pm 30$ μM) without cooperativity.

## Thermodynamics of Hfq-CTR binding

The differential power delivered by the calorimeter, corresponding to the time-stepped titration of 0.5 mM Hfq–CTR with a 2.5 mM (bp) DNA solution in T–buffer ($p$H 7.5), is displayed in Figure 7. The thermogram exhibits a negative signal, which indicates that the reaction is exothermic and associated with heat release. At long times, heat exchanges close to zero mark the decrease of thermodynamic interactions between DNA and Hfq–CTR. The light grey curve in panel A of Figure 7 represents the thermogram obtained from the dilution of DNA in T–buffer. The endothermic signal remains weak and its amplitude is consistent with that of dilution of a polymer or surfactant in this range of concentrations (37, 38). Binding isotherms obtained with different concentrations and molar ratios $X$ of DNA and Hfq–CTR are displayed in panels B–D in Figure 7. Titrations were carried out in various conditions to test the reliability of the assay as well as to explore different ranges of DNA and Hfq–CTR concentrations. The enthalpy curves exhibit a sigmoidal increase with increasing $X$, with initial binding values starting around $-20$ kJ mol$^{-1}$. Until the end of the titration, the enthalpies remain negative, in accordance with an exothermic reaction.

The ITC data were analysed with the Multiple Non-Interacting Sites (MNIS) model (37, 39, 40). This model gives an expression for the heat exchange of the form:

$$\Delta H(X,n,r) = \frac{1}{2}\Delta H_b \left(1 + \frac{n-X-r}{((n+X+r)^2 - 4Xn)^{1/2}}\right) \quad (1)$$

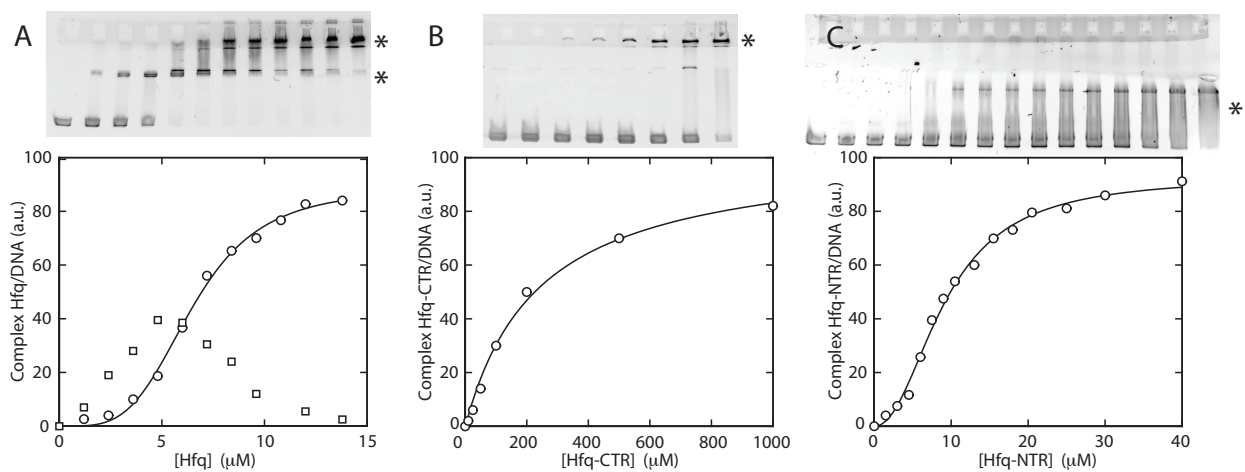

**Figure 6.** *In vitro* analysis of Hfq protein–DNA binding properties. (**A**) Gel shift showing the binding of wild-type Hfq to DNA. (**B**) Binding of Hfq–CTR. (**C**) Binding of Hfq–NTR. The intensities of the asterisk-marked bands are set out in the graphs versus the concentration of monomeric Hfq, Hfq–CTR, and Hfq–NTR, respectively. Note that a significant supershifted species is observed only in the case of wild-type Hfq. Curves represent non-linear least-squares fits of cooperative (Hfq, Hfq–NTR) and non-cooperative (Hfq–CTR) binding models.

where $\Delta H_b$ is the binding enthalpy, $X$ the molar ratio and $n$ the reaction stoichiometry. The coefficient $r$ is defined as $r = 1/(K_b[\text{Hfq–CTR}])$, where $K_b$ denotes the affinity binding constant. Least square calculations using Equation (**1**) give the continuous curves shown in panels B–D in Figure 7. The agreement between the prediction of the MNIS model and the measured heat exchange is excellent, and the fit procedure provides a consistent set of thermodynamic data (collected in Table 2). By taking the average of the data obtained from the three titration experiments, we obtain for the binding enthalpy $\Delta H_b = -23.7 \pm 0.7$ kJ mol$^{-1}$ and for the binding constant $K_b = (2.3 \pm 0.7) \times 10^4$ M$^{-1}$. The averaged value of the stoichiometry parameter $n = 0.6 \pm 0.2$, which implies that approximately two Hfq–CTR peptides interact with one base pair. Knowing the binding enthalpy and binding constant, the changes in reaction free energy $\Delta G$ and entropy $\Delta S$ can be calculated using $\Delta G = -RT \ln K_b$ and $\Delta S = (\Delta H_b - \Delta G)/T$. Again, the data obtained from the three titration assays are consistent, with averaged free energy and entropy changes of $\Delta G = -24.8 \pm 0.8$ kJ mol$^{-1}$ and $\Delta S = 3.7 \pm 0.6$ J mol$^{-1}$ K$^{-1}$, respectively.

## DISCUSSION

Qualitatively, the same compaction behaviour is observed for Hfq–CTR and full length Hfq in the channel systems as well as the bulk phase (28). Condensation in the bulk phase requires, however, more than an order of magnitude higher concentration of protein or peptide. Similar behaviour has previously been reported for another NAP (H-NS), like-charged proteins (bovine serum albumin and hemoglobin) and neutral crowders (34, 41, 42, 43). This shows that the conformation, folding, and condensation of DNA are not only related to classical controlling factors, such as osmotic pressure, charge and ligand binding, but that the interplay with confinement is of paramount importance. The critical (monomer) concentrations of Hfq for condensation are similar to those pertaining to Hfq–CTR. The decrease in stretch in the channel systems and compaction into a condensed form indicate Hfq–CTR mediated bridging interaction between different segments of the DNA molecule. In contrast, the marginal effect of Hfq–NTR on the conformation shows that the hexameric torus is not directly involved in the compaction of DNA. Electrostatics plays a minor role in Hfq-mediated compaction, because similar behaviour is observed for DNA in T–buffer and T–buffer with 30 mM KGlu.

Bridging and looping of longer DNAs by Hfq–CTR is confirmed by the atomic force microscopy images. Furthermore, the images of shorter molecules show a ∼15% decrease in DNA contour length and ∼40% increase in persistence length following binding of Hfq–CTR. These changes in mechanical properties of the double stranded DNA molecule are similar to those previously observed for full length Hfq. However, in order to obtain this result, the Hfq–CTR to bp ratio needed to be increased ten-fold with respect to Hfq, in accordance with Hfq–CTR's relatively low affinity for binding to the double helix as shown by the ITC assay. Note that the changes in contour and persistence length have a minor effect on the extension of the DNA molecules inside the channels. The gradual shortening prior to the collapse into the condensed state is mainly caused by bridging interaction mediated by Hfq–CTR.

The electrophoretic mobility assays show two band shifts for full length Hfq. As previously reported, the high affinity dissociation constant $K_d$ corresponds to binding of the protein to DNA (10). The second band shift with lower affinity and with a migration distance close to the loading wells indicates the formation of complexes of relatively large molecular weight through protein-mediated bridging of multiple DNA molecules. The bridges can be formed by a single protein and/or self-interactions among multiple proteins. For Hfq–CTR and Hfq–NTR, we observed single band shifts with widely different dissociation constants. In the case of Hfq–NTR, the band shift corresponds to binding of the torus with an affinity comparable to the one of full length Hfq. No aggregated complexes are observed, which agrees with the absence of bridging interaction. A surprising result is that no intermediate band shift is observed for Hfq-CTR. This might be related to a marginal effect on the electrophoretic mobility of DNA following binding of a peptide with relatively small molecular weight. However, the band shift close to the wells shows significant bridging and aggregation of DNA at relatively high concentrations of Hfq-CTR. These results indicate that full length Hfq binds on DNA by interfacing its torus, possibly through electrostatic interaction between the phosphates and positively charged residues located at the rim and/or distal face of the torus. However, binding of the torus on the double helix by itself does not result in aggregation. For the assembly of multiple DNA molecules or segments thereof, the CTR is clearly required.

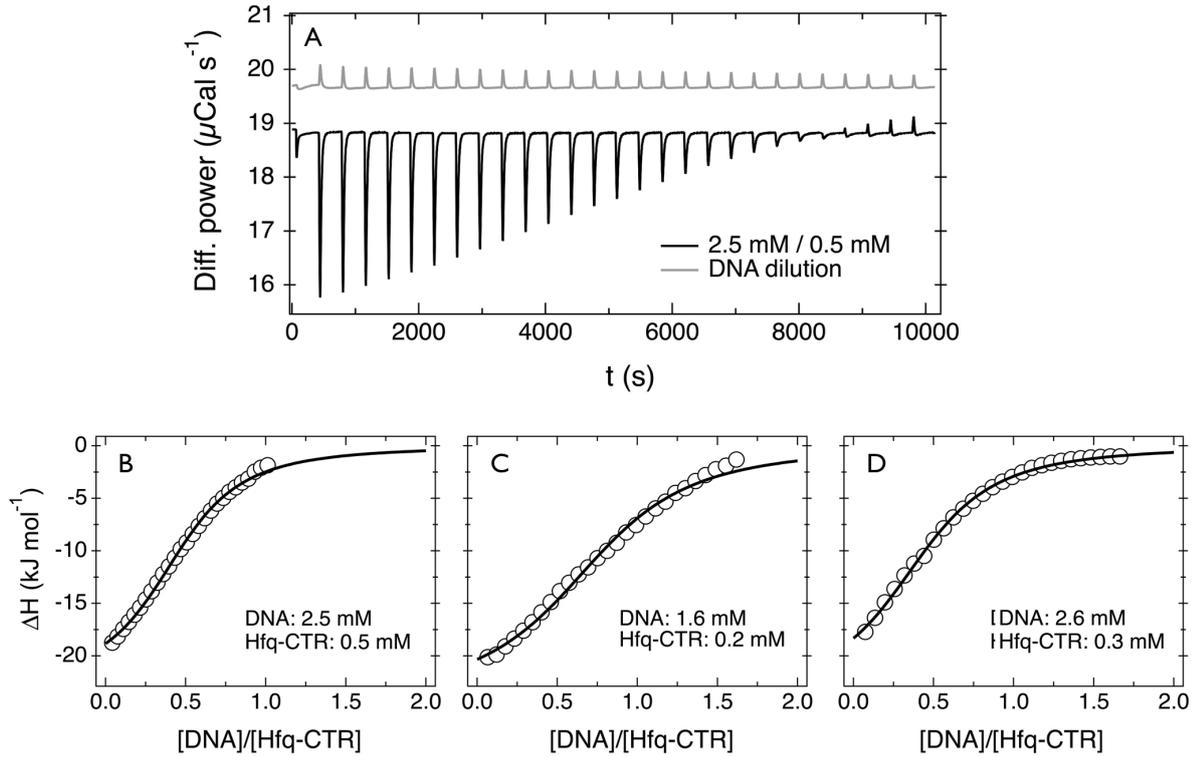

**Figure 7.** (**A**) Differential power as a function of time obtained from calorimetry experiments by titrating Hfq–CTR with DNA (T = 298 K). Molar concentrations are [DNA] = 2.5 mM (bp) and [Hfq–CTR] = 0.5 mM. The upper light-grey curve corresponds to a dilution of DNA in T–buffer. (**B–D**) Binding isotherms obtained in DNA/Hfq–CTR titration experiments performed in different conditions. In all three cases, the enthalpies are negative and the reactions exothermic. The continuous curves represent Equation (**1**) with optimised thermodynamic parameters collected in Table 2.

**Table 2.** Thermodynamic parameters obtained from the fit of Equation (**1**) to the binding isotherms in Figure 7. $\Delta H_b$, $K_b$, $n$, $\Delta G$ and $\Delta S$ denote the binding enthalpy, binding constant, stoichiometry, free energy and entropy changes, respectively.

| [DNA] (mM) | [Hfq–CTR] (mM) | $\Delta H_b$ (kJ mol$^{-1}$) | $K_b$ ($10^4$ M$^{-1}$) | $n$ | $\Delta G$ (kJ mol$^{-1}$) | $\Delta S$ (mol$^{-1}$ K$^{-1}$) |
|---|---|---|---|---|---|---|
| 2.6 | 0.3 | -23.9±0.6 | 2.2±0.1 | 0.51±0.01 | -24.7±0.2 | 3±3 |
| 2.5 | 0.5 | -23.0±0.5 | 1.7±0.1 | 0.51±0.01 | -24.1±0.2 | 4±2 |
| 1.6 | 0.2 | -24.3±0.7 | 3.1±0.3 | 0.81±0.02 | -25.6±0.3 | 4±3 |

The single-molecule compaction, electrophoretic mobility and calorimetry experiments provide information about the propensity for binding and bridging of DNA by Hfq–CTR. Note that the compaction and electrophoretic mobility experiments are sensitive to Hfq–CTR - DNA interaction in an indirect way through the effect on DNA conformation and DNA-DNA aggregation, respectively. In the calorimetry experiment, binding is directly monitored through the exchange of heat. The compaction experiments show a significant effect of nano-confinement, but in the case of the bulk phase assay we obtained a critical Hfq–CTR concentration of around 1 $\mu$M. As inferred from electrophoresis, the dissociation constant associated with DNA aggregation takes a value $K_d = 240\pm30$ $\mu$M. In the calorimetry experiment, the dissociation constant is the inverse of the reaction binding constant $K_b$, so that we obtain $K_d = 1/K_b = 40\pm10$ $\mu$M. This variety in affinity pertaining to different experimental approaches shows that compaction and aggregation by bridging interaction requires different degrees of saturation. A relatively small number of bridges is sufficient for intramolecular compaction, whereas for aggregation of multiple DNA molecules an excess of the bridging ligand is needed. Besides ligand binding, other environmental factors such as DNA concentration, confinement (gel matrix), crowding and buffer conditions are clearly of importance in controlling conformation and intermolecular organisation.

The fitted stoichiometry ratio $n$ implies about two bound Hfq–CTR peptides per base pair. However, not all bound peptides are necessarily anchored to the double helix. They may also be bound by peptide aggregation with concomitant less peptide-DNA interaction. Expressed in thermal energy units, the binding enthalpy, free energy and entropy contributions are around $-9.5$, $-10$, and $0.5$ $k_BT$, respectively. These values agree with a strong, enthalpy driven assembly and a marginal entropic contribution to the reaction thermodynamics. From a comparison with literature data for condensation of DNA by multivalent ions or phospholipids, it follows that the binding of Hfq–CTR on DNA is not controlled by electrostatics (44, 45). In the case of electrostatic complexation, the entropic contribution is expected to be much larger than the enthalpic one, due to the release of counterions associated with ion pairing. Other calorimetric studies on complexation involving nanoparticles, polymers and/or surfactants have shown similar behaviour (37, 38, 46). Furthermore, at physiological $p$H the CTR is almost electroneutral (one basic and four acidic residues), highly flexible, and contains many residues with short polar side-groups. Accordingly, other types of interaction, such as hydrogen-bonding with the relatively abundant histidine and serine residues, must be involved in the binding of the CTR on DNA.

## CONCLUSIONS

From the similar equilibrium dissociation constants of Hfq and Hfq–NTR, it is clear that the torus facilitates binding of full length Hfq on DNA. However, as shown by the nanofludics and bulk phase condensation experiments, binding by itself does not result in compaction of DNA into a condensed state. For compaction, the presence of the CTR is clearly required. Binding of Hfq–CTR results in similar changes in the mechanical properties of the duplex as Hfq. However, as inferred from the moderate increase in bending rigidity, no rigid filament is formed as for another nucleoid associated protein H-NS (41, 47, 48). The decrease in contour length indicates a modification of DNA's secondary structure, that is opening of AT-rich tracts and/or a transition of the duplex from the B- to the A-form (10). Precisely, the transition from the B- to A-form is favoured by low hydration, so that the expulsion of interfacial water by protein binding affects DNA helicity.

In consideration of the present and previous results, we arrive at the following model for binding and bridging of Hfq on the double helix. With high affinity, the hexameric protein lands on the duplex through interfacing its torus, followed by anchoring through binding of one or more of its C-terminal arms. The initial binding is possibly facilitated by electrostatic interaction between the phosphates of the double helix and positively charged residues located at the rim and/or distal face of the torus. Anchoring by binding of the CTR is however strongly enthalpy driven and stabilised by non-electrostatic interactions such as hydrogen bonding with the polar residues. It could also be sequence specific, since structural modification of the duplex is involved as shown by changes in bending rigidity and contour length of the double helix. Once bound, Hfq can form a bridge through anchoring one or more of its other CTR arms on another section of the same or another DNA molecule and/or by CTR-mediated self-interactions among multiple proteins. The propensity for bridging and compaction of DNA by the C-terminal domain might be related to aggregation of bound protein and may have implications for protein binding related gene regulation.

## FUNDING

This work has been supported by NUS-USPC collaboration grant and Singapore Ministry of Education grant MOE2014-T2-1-001. The French Embassy in Poland is gratefully acknowledged for financial support (scholarship to K. K.). Funding for open access charge: Singapore Ministry of Education.

*Conflict of interest statement.* None declared.

## REFERENCES

1. Sun,X., Zhulin,I., and Wartell,R.M. (2002) Predicted structure and phyletic distribution of the RNA-binding protein Hfq. *Nucleic Acids Res.*, **30**, 3662–3671.
2. de Fernandez,M.T.F., Hayward,W.S., and August,J.T. (1972) Bacterial proteins required for replication of phage Q$\beta$ ribonucleic acid. *J. Biol. Chem.*, **247**, 824–831.
3. Tsui,H.C.T., Leung,H.C.E., and Winkler,M.E. (1994) Characterization of broadly pleiotropic phenotypes caused by an hfq insertion mutation in Escherichia coli K-12. *Mol. Microbiol.*, **13**, 35–49.
4. Vogel,J. and Luisi,B.F. (2011) Hfq and its constellation of RNA. *Nature Rev. Microbiol.*, **9**, 578–589.
5. Sobrero,P. and Valverde,C. (2012) The bacterial protein Hfq: much more than a mere RNA-binding factor. *Crit. Rev. Microbiol.*, **38**, 1–24.
6. Gottesman,S. and Storz,G. (2015) RNA reflections: converging on Hfq. *RNA*, **21**, 511–512.
7. Cech,G.M., Pakuła,B., Kamrowska,D., Węgrzyn,G., Arluison,V., and Szalewska-Pałasz,A. (2014) Hfq protein deficiency in Escherichia coli affects ColE1-like but not $\lambda$ plasmid DNA replication. *Plasmid*, **73**, 10–15.
8. Ellis,M.J. and Haniford,D.B. (2016) Riboregulation of bacterial and archaeal transposition. *Wiley Interdisciplinary Reviews: RNA*, **7**, 382–398.
9. Le Derout,J., Boni,I.V., Régnier,P., and Hajnsdorf,E. (2010) Hfq affects mRNA levels independently of degradation. *BMC Mol. Biol.*, **11**, 17.
10. Geinguenaud,F., Calandrini,V., Teixeira,J., Mayer,C., Liquier,J., Lavelle,C., and Arluison,V. (2011) Conformational

transition of DNA bound to Hfq probed by infrared spectroscopy. *Phys. Chem. Chem. Phys.,* **13**, 1222–1229.
11. Sukhodolets,M.V. and Garges,S. (2003) Interaction of Escherichia coli RNA polymerase with the ribosomal protein S1 and the Sm-like ATPase Hfq. *Biochem.,* **42**, 8022–8034.
12. Rabhi,M., Espéli,O., Schwartz,A., Cayrol,B., Rahmouni,A.R., Arluison,V., and Boudvillain,M. (2011) The Sm-like RNA chaperone Hfq mediates transcription antitermination at Rho-dependent terminators. *EMBO J.,* **30**, 2805–2816.
13. Azam,T.A. and Ishihama,A. (1999) Twelve species of the nucleoid-associated protein from Escherichia coli sequence recognition specificity and DNA binding affinity. *J. Biol. Chem.,* **274**, 33105–33113.
14. Azam,T.A., Iwata,A., Nishimura,A., Ueda,S., and Ishihama,A. (1999) Growth phase-dependent variation in protein composition of the Escherichia coli nucleoid. *J. Bacteriology,* **181**, 6361–6370.
15. Azam,T.A., Hiraga,S., and Ishihama,A. (2000) Two types of localization of the DNA-binding proteins within the Escherichia coli nucleoid. *Genes to Cells,* **5**, 613–626.
16. Diestra,E., Cayrol,B., Arluison,V., and Risco,C. (2009) Cellular electron microscopy imaging reveals the localization of the Hfq protein close to the bacterial membrane. *PloS one,* **4**, e8301.
17. Tsui,H.C.T., Feng,G., and Winkler,M.E. (1997) Negative regulation of mutS and mutH repair gene expression by the Hfq and RpoS global regulators of Escherichia coli K-12. *J. Bacteriology,* **179**, 7476–7487.
18. Azam,T.A. and Ishihama,A. (2015) Growth phase dependent changes in the structure and protein composition of nucleoid in Escherichia coli. *Sci. China Life Sci.,* **58**, 902–911.
19. Folichon,M., Arluison,V., Pellegrini,O., Huntzinger,E., Régnier,P., and Hajnsdorf,E. (2003) The poly (A) binding protein Hfq protects RNA from RNase E and exoribonucleolytic degradation. *Nucleic Acids Res.,* **31**, 7302–7310.
20. Kajitani,M. and Ishihama,A. (1991) Identification and sequence determination of the host factor gene for bacteriophage Q$\beta$. *Nucleic Acids Res.,* **19**, 1063–1066.
21. Ohniwa,R.L., Muchaku,H., Saito,S., Wada,C., and Morikawa,K. (2013) Atomic force microscopy analysis of the role of major DNA-binding proteins in organization of the nucleoid in Escherichia coli. *PloS one,* **8**, e72954.
22. Dorman,C.J. (2009) Nucleoid-associated proteins and bacterial physiology. *Adv. Appl. Microbiol.,* **67**, 47–64.
23. Link,T.M., Valentin-Hansen,P., and Brennan,R.G. (2009) Structure of Escherichia coli Hfq bound to polyriboadenylate RNA. *Proc. Natl. Acad. Sci. U.S.A.,* **106**, 19292–19297.
24. Updegrove,T.B., Correia,J.J., Galletto,R., Bujalowski,W., and Wartell,R.M. (2010) E. coli DNA associated with isolated Hfq interacts with Hfq's distal surface and C-terminal domain. *BBA – Gene Regul. Mech.,* **1799**, 588–596.
25. Arluison,V., Folichon,M., Marco,S., Derreumaux,P., Pellegrini,O., Seguin,J., Hajnsdorf,E., and Regnier,P. (2004) The C-terminal domain of Escherichia coli Hfq increases the stability of the hexamer. *Eur. J. Biochem.,* **271**, 1258–1265.
26. Takada,A., Wachi,M., Kaidow,A., Takamura,M., and Nagai,K. (1997) DNA Binding Properties of the hfq Gene Product of Escherichia coli. *Biochem. Biophys. Res. Commun.,* **236**, 576–579.
27. Fortas,E., Piccirilli,F., Malabirade,A., Militello,V., Trépout,S., Marco,S., Taghbalout,A., and Arluison,V. (2015) New insight into the structure and function of Hfq C-terminus. *Bioscience Rep.,* **35**, 1–9.
28. Jiang,K., Zhang,C., Guttula,D., Liu,F., van Kan,J.A., Lavelle,C., Kubiak,K., Malabirade,A., Lapp,A., Arluison,V., and van der Maarel,J.R.C. (2015) Effects of Hfq on the conformation and compaction of DNA. *Nucleic Acids Res.,* **43**, 4332–4341.
29. Zhang,C., Zhang,F., van Kan,J.A., and van der Maarel,J.R.C. (2008) Effects of electrostatic screening on the conformation of single DNA molecules confined in a nanochannel. *J. Chem. Phys.,* **128**, 225109.
30. van Kan,J.A., Zhang,C., Perumal Malar,P., and van der Maarel,J.R.C. (2012) High throughput fabrication of disposable nanofluidic lab-on-chip devices for single molecule studies. *Biomicrofluidics,* **6**, 036502.
31. van Kan,J.A., Bettiol,A.A., and Watt,F. (2006) Proton beam writing of three-dimensional nanostructures in hydrogen silsesquioxane. *Nano Lett.,* **6**, 579–582.
32. van Kan,J.A., Shao,P.G., Wang,Y.H., and Malar,P. (2011) Proton beam writing a platform technology for high quality three-dimensional metal mold fabrication for nanofluidic applications. *Microsyst. Technol.,* **17**, 1519–1527.
33. Johansen,F. and Jacobsen,J.P. (1998) 1H NMR studies of the bis-intercalation of a homodimeric oxazole yellow dye in DNA oligonucleotides. *J. Biomol. Struct. Dyn.,* **16**, 205–222.
34. Zhang,C., Shao,P.G., van Kan,J.A., and van der Maarel,J.R.C. (2009) Macromolecular crowding induced elongation and compaction of single DNA molecules confined in a nanochannel. *Proc. Natl. Acad. Sci. U.S.A.,* **106**, 16651–16656.
35. Wiggins,P.A., van der Heijden,T., Moreno-Herrero,F., Spakowitz,A., Phillips,R., Widom,J., Dekker,C., and Nelson,P.C. (2006) High flexibility of DNA on short length scales probed by atomic force microscopy. *Nat. Nanotechnol.,* **1**, 137–141.
36. Kundukad,B., Cong,P., van der Maarel,J.R.C., and Doyle,P.S. (2013) Time-dependent bending rigidity and helical twist of DNA by rearrangement of bound HU protein. *Nucleic Acids Res.,* **41**, 8280–8288.
37. Courtois,J. and Berret,J.F. (2010) Probing oppositely charged surfactant and copolymer interactions by isothermal titration microcalorimetry. *Langmuir,* **26**, 11750–11758.
38. Vitorazi,L., Ould-Moussa,N., Sekar,S., Fresnais,J., Loh,W., Chapel,J.P., and Berret,J.F. (2014) Evidence of a two-step process and pathway dependency in the thermodynamics of poly (diallyldimethylammonium chloride)/poly (sodium acrylate) complexation. *Soft Matter,* **10**, 9496–9505.
39. Herrera,I. and Winnik,M.A. (2013) Differential binding models for isothermal titration calorimetry: moving beyond the Wiseman isotherm. *J. Phys. Chem. B,* **117**, 8659–8672.
40. Wiseman,T., Williston,S., Brandts,J.F., and Lin,L.N. (1989) Rapid measurement of binding constants and heats of binding using a new titration calorimeter. *Anal. Biochem.,* **179**, 131–137.
41. Zhang,C., Guttula,D., Liu,F., Malar,P.P., Ng,S.Y., Dai,L., Doyle,P.S., van Kan,J.A., and van der Maarel,J.R.C. (2013) Effect of H-NS on the elongation and compaction of single DNA molecules in a nanospace. *Soft Matter,* **9**, 9593–9601.
42. Zhang,C., Gong,Z., Guttula,D., Malar,P.P., van Kan,J.a., Doyle,P.S., and van der Maarel,J.R.C. (2012) Nanouidic compaction of DNA by like-charged protein. *J. Phys. Chem. B,* **116**, 3031–3036.
43. van der Maarel,J.R.C., Zhang,C., and van Kan,J.A. (2014) A nanochannel platform for single DNA studies: from crowding, protein DNA interaction, to sequencing of genomic information. *Isr. J. Chem.,* **54**, 1573–1588.
44. Matulis,D., Rouzina,I., and Bloomfield,V.A. (2000) Thermodynamics of DNA binding and condensation: isothermal titration calorimetry and electrostatic mechanism. *J. Mol. Biol.,* **296**, 1053–1063.
45. Kennedy,M.T., Pozharski,E.V., Rakhmanova,V.A., and MacDonald,R.C. (2000) Factors governing the assembly of cationic phospholipid-DNA complexes. *Biophys. J.,* **78**, 1620–1633.
46. Loosli,F., Vitorazi,L., Berret,J.F., and Stoll,S. (2015) Towards a better understanding on agglomeration mechanisms and thermodynamic properties of TiO 2 nanoparticles interacting with natural organic matter. *Water Research,* **80**, 139–148.
47. Amit,R., Oppenheim,A.B., and Stavans,J. (2003) Increased bending rigidity of single DNA molecules by H-NS, a temperature and osmolarity sensor. *Biophys. J.,* **84**, 2467–2473.
48. van der Maarel,J.R.C., Guttula,D., Arluison,V., Egelhaaf,S.U., Grillo,I., and Forsyth,V.T. (2016) Structure of the H-NS-DNA nucleoprotein complex. *Soft Matter,* **12**, 3636–3642.